# Experimental demonstration of a space-time modulated airborne acoustic circulator

Tinggui Chen[1,2], Matthieu Malléjac[2], Chuanxing Bi[1], Baizhan Xia[3] & Romain Fleury[2*]

*1 Institute of Sound and Vibration Research, Hefei University of Technology, Hefei 230009, China*
*2 Laboratory of Wave Engineering, School of Electrical Engineering, EPFL, Lausanne, Switzerland*
*3 State Key Laboratory of Advanced Design and Manufacturing Technology for Vehicle, Hunan University, Changsha 410082, China*

Achieving strongly nonreciprocal scattering in compact linear acoustic devices is a challenging task. One possible solution is the use of time-modulated resonators, however, their implementation in the realm of audible airborne acoustics is typically hindered by the difficulty to obtain large modulation depth and speeds while managing noise issues. Here, we propose a practical and cost-efficient route to realize simple modulated resonators and observe experimentally the strong nonreciprocal behavior of an acoustic circulator. We propose to modulate the neck cross-section areas of three coupled Helmholtz resonators using rotating circular plates actuated by an electrical motor, and control their phase difference via meshed gears, thereby implementing a modulation scheme with broken time-reversal symmetry that effectively imparts angular momentum to the system. We experimentally demonstrate tunable nonreciprocal behavior with a high nonreciprocal isolation of 34 dB and reflection as low as -9 dB, with insertion losses of 5 dB and parasitic signals below -20 dB. All the experimental results agree well with theoretical and numerical predictions.

[*]romain.fleury@epfl.ch

## I. INTRODUCTION

Rayleigh's reciprocity theorem states that the transmission between two points in linear, time-invariant media remains the same even if the source and receiver positions are interchanged [1-4]. Such reciprocity can be a hindrance in scenarios where unidirectional wave motion is desirable. For instance, one-way acoustic devices, two-port isolators and three-port circulators, require breaking reciprocity. Nonreciprocity acoustic response in magneto-elastic material has been explored [5], but the inherently weak magneto-acoustic coupling effect remains challenging to observe large nonreciprocity as commonly done in the electromagnetic domain using ferromagnetic materials. The two main known strategies for realizing acoustic nonreciprocity effects are either based on nonlinearity [6-9], or on linear kinetic media, for example fluids with directional flow [10-13]. Nonreciprocity with nonlinear media suffers from signal distortion, requires high energy intensity, and is plagued by dynamic reciprocity constraints [14]. In contrast, the introduction of a directional flow in the system can overcome these issues and allows both linear [10] and nonlinear [15] nonreciprocity.

An interesting alternative to induce linear acoustic nonreciprocity, spatiotemporal modulation, has attracted a great deal of interest over the past decade [16-24]. With such modulation, an effective angular momentum can be imparted to an acoustic system by breaking time-reversal symmetry and therefore induce strong nonreciprocal effects. Based on this idea, some research has recently been devoted to magnetic-free nonreciprocal acoustic propagation, such as proposals for parametric amplification [16-19], frequency conversion [20-22], Floquet topological insulators [23,24], etc., but most of them stay at the theoretical and numerical investigations level. Only a few experimental demonstrations focusing on two port isolators have been reported so far in the realm of airborne acoustics at audible frequencies. Among others, coupled resonators were mechanically modulated by shakers [25] or with synthetic effective magnetic bias obtained through electrically modulated coupling phase [26]. The design of a circulator, on the other hand, has a higher level of complexity. The theory for subwavelength circulators based on three time-modulated resonators was proposed in 2015 [27]. We note recent experimental investigations of this technique in the elastodynamic regime, using modulated piezoelectric patches [28], or in the audible sound regime but with active control techniques applied to modulate loudspeaker dynamics [29]. A simple mechanical modulation scheme with sufficient modulation speed, depth, and low-noise is still missing and would nicely complement the toolbox of available methods to create airborne audible acoustic circulators.

We unveil in this work a practical and cost-efficient route to demonstrate experimentally the strong nonreciprocal behavior for sound circulation in a three-port network. The system consists of three acoustic cavities coupled by neck channels, the cross-section areas of which are periodically modulated in time, by mechanical means, at a much lower frequency than the incident sound. By doing so, the resonance frequencies of each individual resonator can be modulated dynamically in a rotating fashion, allowing to impart a momentum bias to the system, which is ultimately responsible for the acoustic circulation. We show that a careful design of the circulator geometry leads to good performances. By slightly changing the external coupling parameters in experiment, the highest possible isolation can reach 34 dB whereas the lowest reflection is -9 dB, with insertion losses of 5 dB and parasitic signals below -20 dB.

## II. EXPERIMENTAL SETUP

Inspired by the three-port network theory proposed in a prior art [27], our design consists in three cylindrical air-filled acoustic cavities (diameter $d$, height $h$), coupled to each other via internal coupling channels (diameter $d_2$, length $l_2$), referred as the necks in the following, forming an acoustic resonator with 120° rotational symmetry (see Fig. 1(a)). Three additional channels (diameter $d_1$, length $l_1$) couple this resonator to external waveguides, defining a three-port scattering system, or network. Instead of modulating the cavity volumes (effective bulk modulus), as in Ref. [27], we break reciprocity by modulating in a rotating fashion the effective cross-section area of three necks (effective density). The cross-section area $A$ of neck 1 is modulated by $\Delta A_1 = A_m \cos(\omega_m t)$, whereas the cross-section areas of necks 2 and 3 are modulated at the same modulation frequency $f_m = \omega_m/2\pi$ and strength $A_m$, but with $2\pi/3$ and $4\pi/3$ phase delays, i.e., $\Delta A_2 = A_m \cos(\omega_m t - 2\pi/3)$ and $\Delta A_3 = A_m \cos(\omega_m t - 4\pi/3)$, respectively. The schematic of the designed experimental setup is shown in Fig. 1 (b). The cross-section modulation is achieved by rotating periodically circular plates (diameter $d_p$, thickness $t_p$) inserted in the necks. By doing so, the neck effective cross-section areas of each individual resonator are changed dynamically. The rotating plates are connected by rods to three meshed gears mutually driven by an electrical motor to guarantee an accurate and constant phase difference. Moreover, the modulation frequency, which is directly proportional to the plate rotation speed inside the necks, can be easily tuned by varying the motor output voltage and current.

To design the setup, both analytical lumped circuit/coupled mode theory modeling (See Appendix A) and full-wave simulation were used. It is worth noting that the theoretical framework can be applied to three-port network with any geometrical parameters. The geometrical parameters of the experimental setup illustrated in Fig. 1(c), were chosen carefully to provide a sufficiently high quality factor (Q), enabling an enhancement of the nonreciprocal behavior by the resonance, required to observe circulation at low modulation speeds and depth. The value of the Q factor is controlled by tuning the external radiation loss (via $d_1$) and is ultimately limited by internal viscothermal losses, which increase with frequency. The resonance frequency, on the other hand, depends on the geometry of the resonator and scales inversely with its size. The designed circulator, whose dimensions are reported in Table 1, has been 3D printed and is connected to external waveguides with a square cross-section (width of 30 mm), avoiding leakages. The three-port system is excited at port 1 by an electrodynamic loudspeaker delivering sinusoidal signals, while anechoic terminations are connected to ports 2 and 3, to facilitate the scattering characterization.

TABLE 1 Geometrical parameters of the experimental setup

| Parameters | $h$ | $d$ | $d_1$ | $l_1$ | $d_2$ | $l_2$ | $d_p$ | $t_p$ |
|---|---|---|---|---|---|---|---|---|
| Values (mm) | 40 | 20 | 10 | 20 | 20 | 20 | 17 | 0.3 |

The scattering parameters of the three-port system, i.e., the reflection at port 1, $|S_{11}|$, the transmission at ports 2 and 3, $|S_{21}|$ and $|S_{31}|$ respectively, are characterized by three sets of two ¼ inch microphones separated by $s = 50$ mm to discriminate between incident and reflected waves on each port (see Appendix A). The distance between the printed sample and closest microphone is $l = 200$ mm. Closed-up photographs of the setup are provided by the left (motor-driven meshed gears) and right (inner circulator view) insets in Fig. 1(c). The motor is powered by a DC supply with tunable voltage and current.

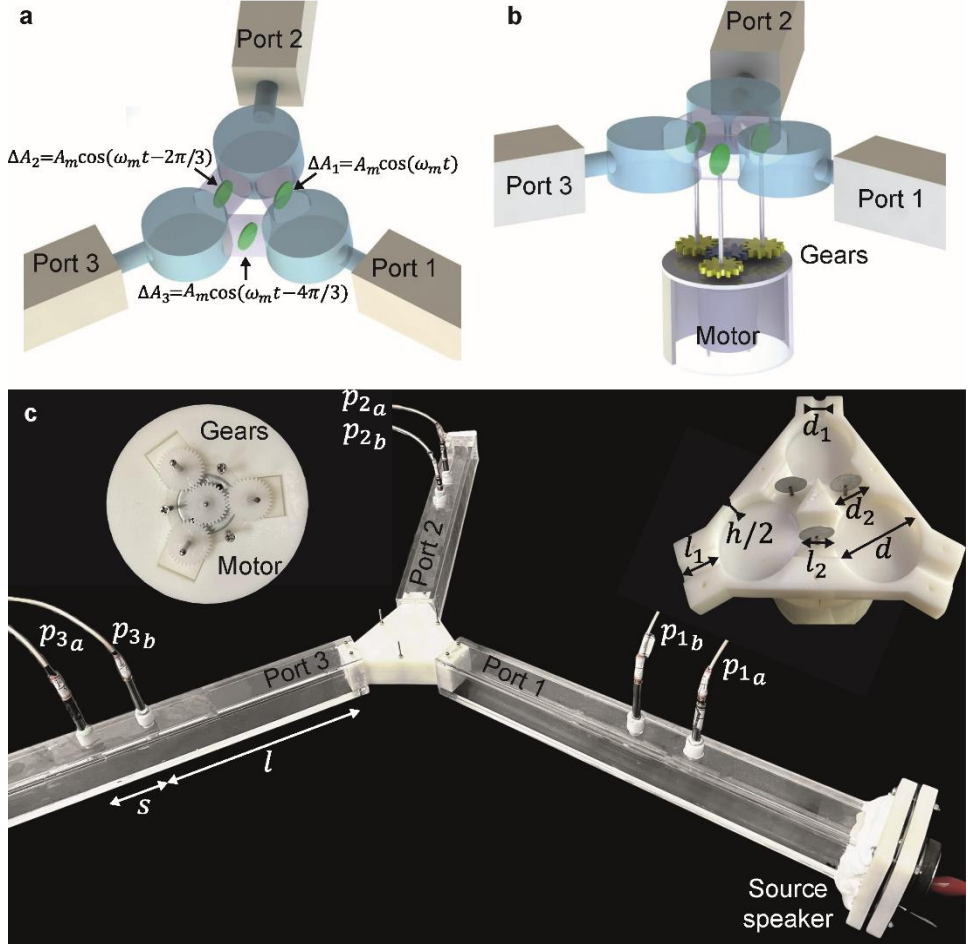

FIG. 1. **Experimental setup:** (a) Schematic of the three-port network composed of three cavities connected via internal coupling channels (necks) and coupled to three external square waveguides. The cross-section areas $A$ of the necks are weakly modulated in a rotating fashion, with amplitude $A_m$ and circular frequency $\omega_m$. (b) Exploded view of the three-port network, highlighting the modulation mechanism. The rotation of the plates is entrained by meshed gears to generate the $2\pi/3$ modulation phase difference between two adjacent necks. (c) Photograph of the experimental setup. A source loudspeaker emits plane incident waves at port 1. The scattering, i.e., the reflection at port 1 and transmissions at ports 2 and 3, are characterized using pair of microphones positioned on each of the external waveguide. Left inset: Photograph of the meshed gears driven by the motor. Right inset: Cut view of the 3D printed three-port resonator.

## III. RESULTS

We performed theoretical, numerical, and experimental investigations to demonstrate the effectiveness of the proposed device. We first consider the time-invariant behavior, for which the three-port network is symmetric and reciprocal. In such configuration, if the system was completely lossless and perfectly three-fold symmetric, the incident energy from port 1 should be equally splitted to ports 2 and 3, with $|S_{21}| = |S_{31}| = 2/3$, and $|S_{11}| = 1/3$ at the resonance frequency of the resonator. Figure 2(a) shows the measured scattering coefficient obtained for the static case. As we can see, although the transmission is equal at ports 2 and 3, the measured transmission amplitude is a bit lower than expected for a lossless system, with $|S_{21}| = |S_{31}| = 0.45$, evidencing the presence of viscothermal losses. To avoid unnecessary return losses and optimal interaction of the incident energy with the resonator, the design has been adjusted to be impedance matched, as confirmed by the relatively low measured reflectance, $|S_{11}| = 0.02$. The time-invariant resonator therefore behaves like an energy divider, as corroborated with the pressure distribution shown in the inset of Fig. 2(a), resulting from the finite element modeling of the whole experimental setup (details on the modeling can be found in Appendix A).

When the motor is operating, the out-of-phase rotations of the plates induce a synthetic angular momentum bias, lifting the degeneracy of the two azimuthal counter-propagating modes in the resonator, and resulting in broken time-reversal symmetry and strong nonreciprocal behavior. As a result, the transmission at ports 2 and 3 are no longer equal. The level of nonreciprocity is sensitive to both modulation frequency $f_m$ and depth $A_m$. In the proposed design, the modulation depth is directly determined by the change in neck's effective cross-section and thus by the diameter of the rotating plates. To determine the modulation depth experimentally, the resonance frequency of the resonator is measured for two different positions of the plates in the necks, parallel or normal to the cross-section, i.e., maximal and minimal effective area respectively, and compared with the mean frequency $\overline{f_r}$. We found $A_m = \frac{\Delta f_r}{\overline{f_r}} = 0.09$ (where $\Delta f_r$ is the difference of both resonance frequency between the two angle of the plate). With $A_m = 9\%$ being fixed, the level of non-reciprocity can be fine-tuned with the second modulation parameter, namely the modulation frequency $f_m$. To better understand the effect of modulation frequency on circulation performance, Fig. 2(b) shows the numerical and measured scattering parameters at resonance as a function of modulation frequency, in dashed lines and symbols respectively. We observe that a maximum value for $|S_{31}|$ is achieved when the modulation frequency reaches 102 Hz, while the minimum transmission for $|S_{21}|$ is reached at $f_m = 68$ Hz. For reflection at port 1, $|S_{11}|$, a higher modulation frequency leads to a decrease in $|S_{11}|$, indicating that low reflection can be achieved at the cost of a higher modulation frequency. These behaviors were theoretically predicted in Ref. [27] and are now confirmed experimentally.

Classically, the performances of a circulator are characterized using four main metrics: (i) the isolation, i.e., the ability to transmit in a single direction $IS = 20 \log |S_{31}/S_{13}|$; (ii) the reflection magnitude in logarithmic scale $R = 20 \log |S_{11}|$;

(iii) the insertion loss $IL = -20\log|S_{31}|$, which characterizes the amplitude loss during the transmission; and (iv) the intermodulation strength of the parasitic signals at $\omega \pm \omega_m$, $P = 20\log|S_{31}^{\omega+\omega_m}|$. Here the isolation in logarithmic is modified as $IS = 20\log|S_{31}/S_{21}|$ due to the symmetry, allowing to measure it from a single port measurement. The ideal circulator should have the highest possible isolation with the lowest reflection, insertion loss, and parasitic signals. Since here the insertion loss, with a few decibels, is insensitive to the modulation frequency and the parasitic signals are always below -18 dB, we only investigate the evolution of the isolation and reflection versus the modulation frequency, shown in Fig. 2(c), in red (left axis) and blue (right axis) respectively. Although the maximum isolation reaches 17 dB at $f_m = 82$ Hz, the best design also requires a compromise with the reflection magnitude, which must be as low as possible. Trading off a bit of reflection, we chose $f_m = 93$ Hz to observe circulation behavior.

Figure 2(d) shows the scattering parameters obtained with a modulation frequency $f_m = 93$ Hz. The amplitude transmission coefficient $|S_{31}|$ increased from 0.45 to 0.56 while $|S_{21}|$ decreased from 0.45 to 0.08 compared to the static configuration. The maximum transmission contrast is observed at 1767 Hz, where resonance takes place, evidencing the circulation behavior. Meanwhile, the minimum reflection coefficient $|S_{11}|$ goes up since the impedance matching condition is no longer met with the modulation. It is worth noting that the measured resonance frequency in the modulated case is changed from 1832 Hz to 1767 Hz since the plates, rods connecting the plates, and the complex flux occurring with the rotation result in slight changes in geometry and additional losses. To compensate for this frequency shift, we slightly increase the volume of the cavities in the numerical model, which consequently downshifts the resonance frequency. Compared with the unbiased resonator, the imparted angular momentum produces an asymmetric field distribution and circulation shown in the inset of Fig. 2(d). The coupled mode theory (CMT) is also used to predict analytically the scattering parameters. Since the loss in experiment is inevitable, the total decay rate $\gamma_T$ consists in a combination of radiation loss $\gamma_R$ and dissipation loss $\gamma_D$ (See Appendix A). The decay rate is then fitted to match the experiments, where $\gamma_R = 38.2\pi s^{-1}$ and $\gamma_D = 20.8\pi s^{-1}$. We found an excellent agreement between CMT (solid lines), FEM (dashed lines) and experiment (symbols).

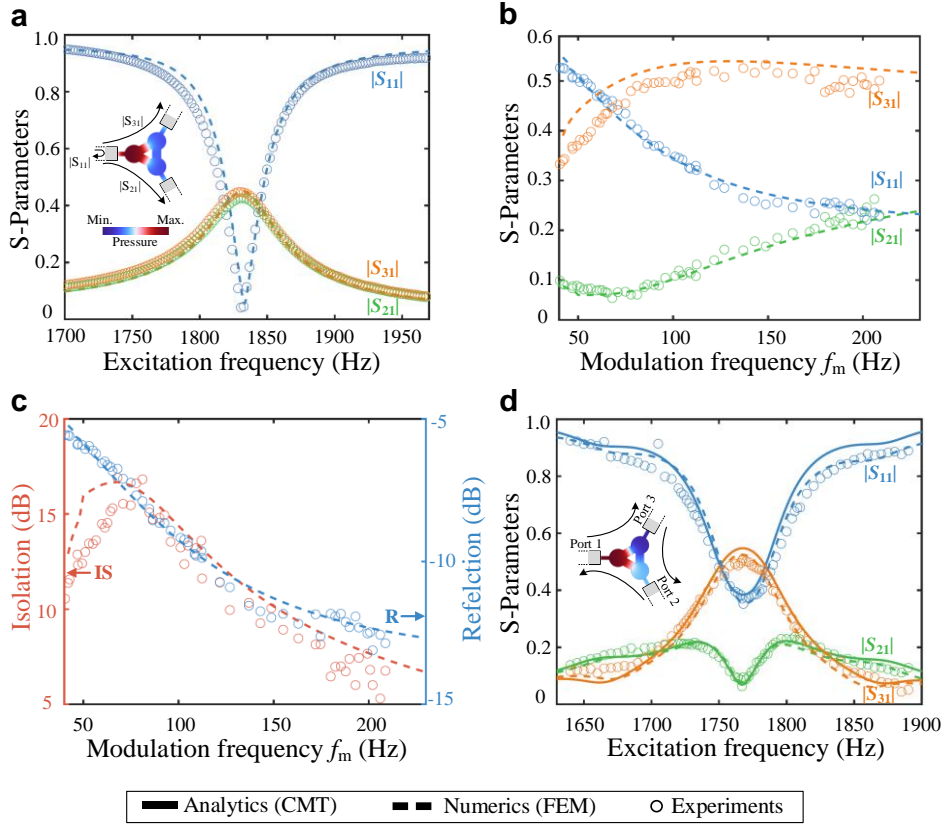

FIG. 2. **Scattering and performance of the circulator in the static and modulated cases**: (a) Reflection at port 1 (blue) and transmission to ports 2 (green) and 3 (orange) in the time-invariant case. The energy incident from port 1 is equally splitted to output ports 2 and 3 simultaneously. (b) Evolution of the scattering parameters with the modulation frequency for a fixed modulation depth. (c) Evolution of the isolation and reflection in logarithmic scale. The maximum isolation can reach 17 dB when $f_m = 82$ Hz while the reflection decreases with increasing modulation frequency. (d) Measured (circle symbols), analytical (solid line) and numerical (dashed line) scattering parameters when the three-port circulator is modulated with a frequency $f_m = 93$ Hz. The incident energy is transmitted to port 3 while blocked at port 2, thus demonstrating a strong nonreciprocity. Excellent agreement between theory (CMT, solid line), simulation (FEM, dashed line) and experiment (circle symbols) is observed.

The proposed time modulated circulator, labeled circulator 1, has been designed to minimize the reflection. Nevertheless, some other applications might require to focus more on the isolation. In such case, one can vary the coupling strength by slightly changing the external coupling channel's diameter. For sake of example, we decreased the external channel diameter from 10 mm to 8 mm, and labeled this second circulator, circulator 2. The decrease of the coupling geometry results in a higher Q factor that enables stronger interaction in the cavities. Figure 3(a) shows the measured scattering parameters with the optimal modulation frequency $f_m = 103$ Hz. Compared with the circulator 1, more energy is reflected back, due to the reduction of cross-section, and a decrease of the measured transmission is observed. In this scenario, the maximum transmission coefficient $|S_{31}| = 0.44$ while the minimum transmission coefficient $|S_{21}|$ goes to a very small value, and a very large isolation of more than 30 dB is achieved. The maximum transmission contrast is observed at 1720 Hz, where resonance takes place. Since the losses changed with this new geometry, the loss parameters have been updated in the CMT model to $\gamma_R = 22\pi s^{-1}$ and $\gamma_D = 22\pi s^{-1}$. In such case ($\gamma_R = \gamma_D$), we get the highest Q factor [30]. A good agreement is observed between the analytics, numerical, and experimental results.

To clearly compare the circulation performance of circulator 1 and 2, Fig. 3(b) displays isolations, reflections, insertion losses, and parasitic signals of the two circulators. Circulator 2's maximum isolation can reach 34 dB at resonance, instead of the 17 dB reachable with circulator 1. Besides, the minimum reflection of circulator 2 is -5 dB, with insertion loss of 7 dB, while the minimum reflection of circulator 1 can reach -9 dB, with insertion loss of 5 dB. Interestingly, the parasitic signals generated by modulation in both two circulators are less than -20 dB. The presented results indicates that the circulation performance can be easily tuned by modifying the modulation frequency and external coupling strength, depending on the targeted application.

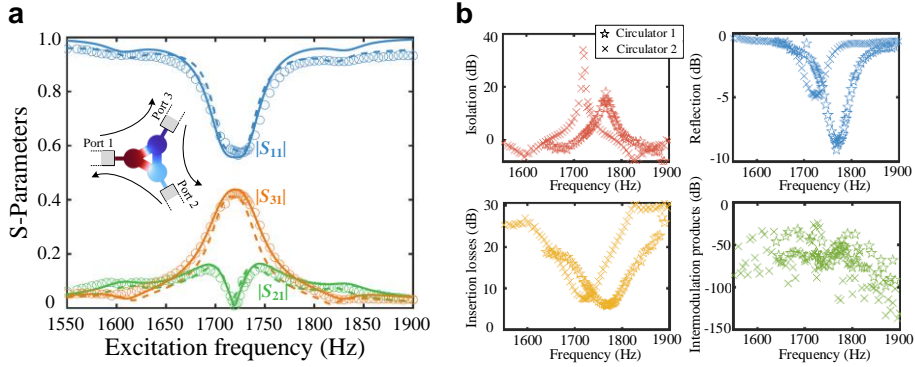

FIG. 3. **Performance comparison for circulators 1 and 2 (for which the diameter of the external coupling channels is changed from 10 mm to 8 mm to improve isolation):** (a) Analytical (solid line), numerical (dashed line) and measured (symbols) scattering parameters for circulator 2 modulated with a frequency $f_m = 103$ Hz. An almost zero transmission is observed at port 2 evidencing the strong nonreciprocity and circulation of the acoustic wave. (b) Evolution of the isolation (red, top left panel), reflection (blue, top right panel), insertion loss (yellow, bottom left panel), and strength of the intermodulation products (green, bottom right panel) with the excitation frequency. For circulator 1, represented by the pentagram symbols, the maximum isolation reaches 17 dB while the minimum reflection reaches -9 dB, with insertion loss of 5 dB at resonance frequency. For circulator 2, represented by the "x" symbols, the maximum isolation reaches 34 dB while the minimum reflection reaches -5 dB, with insertion loss of 7 dB at resonance frequency. The parasitic signals in both two circulators are below -20 dB.

## IV. DISCUSSIONS

The designed experimental setup demonstrates large nonreciprocity in a compact linear three-port resonant acoustic system via spatiotemporal modulation. The circulation behavior is triggered by a low modulation frequency (around 100 Hz), one order of magnitude lower than the resonance frequency (around 1750 Hz), easily achievable with rotating motor. According to Fig. 2, the reflection of the designed circulator can reach low values as long as the modulation frequency is high enough. Besides tuning the voltage and current of the motor, higher modulation frequencies can be reached by rearranging the gears to improve the transmission ratios, or by replacing the motor with a faster one. Ultimately, a too high modulation frequency may introduce more noise, vibration and instabilities that could be detrimental to the system. On the other hand, the modulation depth could be further improved by optimizing the shaped and size of the rotating plate as well as internal waveguide. For example, changing the plate diameter from 17 mm to 18 mm increases the measured modulation depth from 0.09 to 0.11, thus allowing to down shift the plots in Figs. 2(b-c). Similar performance could therefore be achieved with a lower modulation frequency. Additionally, according to the CMT analysis, the higher the Q factor, the lower the required modulation frequency for a given performance, which provides another path to reduce the modulation frequency. The Q factor of our designed circulators can be increased by tailoring external coupling strength. According to Fig. 3, when the external coupling diameter is changed from 10 mm to 8 mm, the value of Q factor is improved from 30 to 39, and as a result, the maximum isolation value is improved from 17 dB to 34 dB. This strong tunning capability is an interesting design feature.

The performance of the current nonreciprocal device is not fundamentally limited, and the geometrical parameters could be further optimized to improve the circulation performance without changing the total size of the system. Once the circulator has been carefully designed, it is possible to find an optimum modulation frequency and depth pair for which the signal entering port 1 is routed exclusively to port 3 and not to port 2. Modulating the coupling between cavities instead of the cavities themselves opens new avenues for the practical implementation of circulators in acoustics. In addition, the proposed modulation strategy is a purely mechanical, magnetless, cost-effective, linear, low-noise system that is easily tunable and scalable.

## V. CONCLUSIONS

To conclude, we have proposed a practical and cost-efficient route in experiment to induce nonreciprocity through an airborne acoustic circulator working at audible frequency based on a mechanical modulation of the effective coupling between cavities. We have evidenced that both high isolation and low reflection can be achieved by adequately tuning the modulation frequency and external coupling strength. The experimental results indicate that the highest isolation can reach 34 dB whereas the lowest reflection is -9 dB, with insertion losses of 5 dB and parasitic signals below -20 dB. The presented results provide a practical modulation strategy for sound isolation, non-Hermitian acoustics, Floquet topological acoustics, among others.

## ACKNOWLEDGEMENTS

The authors thank Dr. Zhe Zhang, Xinxin Guo for insightful discussions. This work was supported by the Fundamental Research Funds for Central Universities

(Grants No. JZ2023HGQA0130 and No. JZ2024HGTA0175).

## APPENDIX A: DETAILS ON THE MODELING
### 1. CMT model and scattering parameters

To analyse the response of the proposed experimental setup, we consider the equivalent lumped circuit model, represented in Fig. 4. The main difference between our lumped circuit model and that of Ref. [27], is that the modulation is performed on the necks' cross-section areas rather than the cavities' volume. As a result, the inductors now vary in time, while the capacitors are time invariant, i.e., a modulation $\Delta L_1$, $\Delta L_2$ and $\Delta L_3$ is applied in a rotating fashion to the three inductors. Furthermore, the acoustic medium filling the cavities, neck, coupling channels, etc. is air (density $\rho_0 = 1.21\ kg/m^3$, sound velocity $c_0 = 343\ m/s$), and dissipation losses are accounted for. As a result, resistances $R_1$, $R_2$ and $R_3$ are introduced in the lumped circuit model.

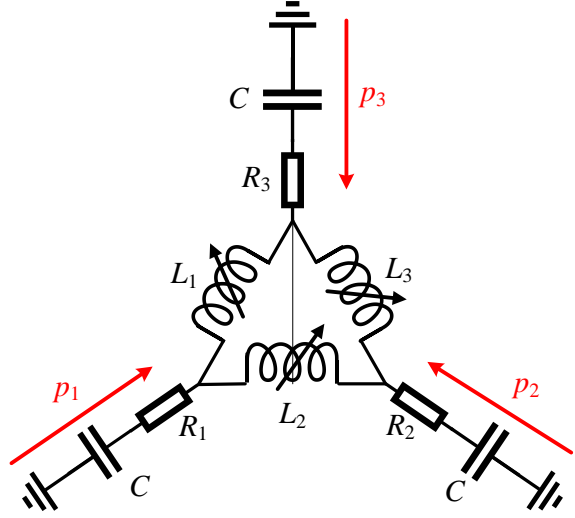

FIG. 4. Lumped elements model of the resonator alone.

To derive the CMT model, we first neglect the dissipation losses. The coupled mode equation derived in Ref. [27], can therefore be simplified to

$$\frac{d^2}{dt^2}a_+ = -\omega_\pm^2 a_+ + \chi e^{i\omega_m t}a_-$$
$$\frac{d^2}{dt^2}a_- = -\omega_\pm^2 a_- + \chi e^{-i\omega_m t}a_+$$
(A1)

where $\alpha_{+(-)}$ is the normalized field amplitude of the counterclockwise (clockwise) rotating mode of the cavity and $\chi = \omega_\pm^2 \delta L/(2L_0)$. $\delta L$ and $L_0$ are modulation depth and acoustic inductance respectively.

To evaluate the scattering properties of the device, the decay rate of each resonator is defined by $\gamma_T$. To account for the losses in the system, the decay rate is composed of both radiation loss $\gamma_R$ and dissipation loss $\gamma_D$. The resonance frequency is thus modified as

$$\Omega_\pm = \omega_\pm + i\gamma_T \tag{A2}$$

Accounting for the 120 degree rotational symmetry and the time-reversal properties, Eq. (A1) can then be rewritten as:

$$\frac{d^2}{dt^2}a_+ = -\Omega_\pm^2 a_+ + \chi e^{i\omega_m t}a_- - 2i\omega_\pm\sqrt{\frac{2\gamma_R}{3}}(S_1^+ + e^{-i\frac{2\pi}{3}}S_2^+ + e^{-i\frac{4\pi}{3}}S_3^+)$$

$$\frac{d^2}{dt^2}a_- = -\Omega_\pm^2 a_- + \chi e^{-i\omega_m t}a_+ - 2i\omega_\pm\sqrt{\frac{2\gamma_R}{3}}(S_1^+ + e^{i\frac{2\pi}{3}}S_2^+ + e^{i\frac{4\pi}{3}}S_3^+) \tag{A3}$$

The normalized amplitude $S_i^-$ of the outgoing wave at port $i$ are given by

$$S_1^- = -S_1^+ + \sqrt{\frac{2\gamma_R}{3}}(a_+ + a_-) \tag{A4}$$

$$S_2^- = -S_2^+ + \sqrt{\frac{2\gamma_R}{3}}(a_+ e^{i\frac{2\pi}{3}} + a_- e^{-i\frac{2\pi}{3}}) \tag{A5}$$

$$S_3^- = -S_3^+ + \sqrt{\frac{2\gamma_R}{3}}(a_+ e^{i\frac{4\pi}{3}} + a_- e^{-i\frac{4\pi}{3}}) \tag{A6}$$

For simplicity, we assume that an incident wave of frequency $\omega$ only excite the system from port 1. As a result, we can write $S_1^+ = e^{i\omega t}$ and $S_2^+ = S_3^+ = 0$, where $S_i^+$ is the normalized amplitude of the incoming wave at port $i$. Eq. (A3) can be further derived as

$$\frac{d^2}{dt^2}a_+ = -\Omega_\pm^2 a_+ + \chi e^{i\omega_m t}a_- - 2i\omega_\pm\sqrt{\frac{2\gamma_R}{3}}e^{-i\omega t}$$

$$\frac{d^2}{dt^2}a_- = -\Omega_\pm^2 a_- + \chi e^{i\omega_m t}a_+ - 2i\omega_\pm\sqrt{\frac{2\gamma_R}{3}}e^{-i\omega t} \tag{A7}$$

It can be seen that Eq. (A7) is satisfied if

$$a_+ = \alpha_+^0 e^{-i\omega t} + \alpha_+^- e^{-i(\omega-\omega_m)t}$$

$$a_- = \alpha_-^0 e^{-i\omega t} + \alpha_-^+ e^{-i(\omega+\omega_m)t} \tag{A8}$$

where the coefficients $\alpha_i^j$ are constant. Then Eq. (A7) becomes

$$\begin{pmatrix} \Omega_\pm^2 - \omega^2 & 0 & 0 & -\chi \\ 0 & \Omega_\pm^2 - \omega^2 & -\chi & 0 \\ 0 & -\chi & \Omega_\pm^2 - (\omega-\omega_m)^2 & 0 \\ -\chi & 0 & 0 & \Omega_\pm^2 - (\omega-\omega_m)^2 \end{pmatrix} \begin{pmatrix} \alpha_+^0 \\ \alpha_-^0 \\ \alpha_+^- \\ \alpha_-^+ \end{pmatrix} = \begin{pmatrix} -2i\omega_\pm\sqrt{2\gamma_R/3} \\ -2i\omega_\pm\sqrt{2\gamma_R/3} \\ 0 \\ 0 \end{pmatrix} \tag{A9}$$

After some algebra, the components $\alpha_+^0$ and $\alpha_-^0$ at $\omega$ can be derived as follows:

$$\alpha_+^0 = -2i\omega_\pm\sqrt{\frac{2\gamma_R}{3}}\frac{\Omega_\pm^2-(\omega+\omega_m)^2}{(\omega^2-\Omega_\pm^2)[(\omega+\omega_m)^2-\Omega_\pm^2]-\chi^2} \tag{A10}$$

$$\alpha_-^0 = -2i\omega_\pm\sqrt{\frac{2\gamma_R}{3}}\frac{\Omega_\pm^2-(\omega-\omega_m)^2}{(\omega^2-\Omega_\pm^2)[(\omega-\omega_m)^2-\Omega_\pm^2]-\chi^2} \tag{A11}$$

The scattering coefficients $S_{ij}^\omega$ at $\omega$ read as

$$S_{11}^\omega = -1 + \sqrt{\frac{2\gamma_R}{3}}(\alpha_+^0 + \alpha_-^0)$$
$$S_{21}^\omega = \sqrt{\frac{2\gamma_R}{3}}(\alpha_+^0 e^{i\frac{2\pi}{3}} + \alpha_-^0 e^{-i\frac{2\pi}{3}}) \quad (A12)$$
$$S_{31}^\omega = \sqrt{\frac{2\gamma_R}{3}}(\alpha_+^0 e^{i\frac{4\pi}{3}} + \alpha_-^0 e^{-i\frac{4\pi}{3}})$$

## 2. Finite-element simulations

To perform a full-wave simulation of the proposed device, we first need to find a way to model the modulated effective cross-sectional area of the neck (modulated inductance by $\delta L/L_0$ in the equivalent lumped model, $L_0 = \rho h_n/S_n$, where $h_n$ is neck height and $S_n$ is neck cross-section area), following the procedure proposed in Ref. [27]. To keep a constant geometry and mesh along the simulation, we account for the modulation by assigning a time-varying density to the air medium filling the resonator. We therefore modify the constitutive equation of the filling medium

$$\nabla p(\mathbf{r},t) = -\frac{d}{dt}[\rho(\mathbf{r},t)\mathbf{u}(\mathbf{r},t)] \quad (A13)$$

$$\nabla \mathbf{u}(\mathbf{r},t) = -\frac{1}{KP_0}\frac{d}{dt}[p(\mathbf{r},t)] \quad (A14)$$

where $p(\mathbf{r},t)$ is the acoustic pressure, $\mathbf{u}(\mathbf{r},t)$ is the particle velocity, and $K$ and $P_0$ are the air bulk modulus and the atmospheric pressure respectively and are time-invariant, and

$$\rho(\mathbf{r},t) = \rho_0 + \delta\rho(\mathbf{r})\cos(\omega_m t - \varphi(\mathbf{r})) \quad (A15)$$

is the dynamically modulated density of the structure. By inserting Eq. (1.14) in Eq. (A13), the obtained wave equation is $\Delta p(\mathbf{r},t) = \frac{1}{KP_0}\frac{d}{dt^2}[\rho(\mathbf{r},t)p(\mathbf{r},t)]$. According to Floquet-Bloch theorem, the acoustic pressure field in the modulated resonator can be written as a Fourier series $p(\mathbf{r},t) = \sum_n f_n(\mathbf{r})e^{i(\omega+n\omega_m)t}$. After some algebra, we obtain an infinite linear set of coupled time-independent differential equations for an arbitrary harmonic of order $n$

$$\Delta f_n(\mathbf{r}) + \frac{1}{KP_0}\rho(\mathbf{r})(\omega+n\omega_m)^2 f_n(\mathbf{r}) = -\frac{1}{2KP_0}\delta\rho(\mathbf{r})(\omega+n\omega_m)^2[f_{n-1}(\mathbf{r})e^{-i\varphi(\mathbf{r})} + f_{n+1}(\mathbf{r})e^{i\varphi(\mathbf{r})}] \quad (A16)$$

Since the modulation frequency is relatively small compared with resonance frequency, we truncate this infinite system to the three harmonics $n = \{-1,0,1\}$ that dominate the field in the frequency range of interest. Thus, the three coupled differential equations obtained by Eq. (A16) can be transformed into weak form and solved in the (quasi-) frequency domain (acoustic module, COMSOL Multiphysics). Furthermore, the narrow region acoustics model is applied to different parts of circulator to evaluate the losses.

## 3. Fitting procedure

We take circulator 1 as an example to give a comprehensive explanation on how the numerical model is fitted to the experimental results.

On the one hand, in the time invariant configuration, the resonance frequency and transmission of the static circulator are 1806 Hz and 0.49 respectively according to the FEM simulation of the geometry reported in Table 1, while the measured values are respectively 1832 Hz and 0.45 in experiment. The difference on the resonance frequency can be explained by fabrication inaccuracy while the difference in transmission value can be associated to additional viscothermal losses due to the 3D printed process. To readjust the numerical model, we slightly decrease the diameter of the cavity from 40 mm to 39.4 mm and add additional losses at the external coupling. The fitting results are presented in Fig. 2(a).

On the other hand, for the time modulated case, the measured resonance frequency 1832 Hz is shifted to 1767 Hz since the plates, rods connecting the plates, and the complex flux occurring with the rotation, result in slight changes in geometry and additional losses. Notably, additional losses, caused by complex flow in the necks, are not properly accounted for in the FEM simulations. The hypothesis is corroborated by the maximum transmission that is lower in the measurement than in the simulation. To account for this additional mismatch, we further modify the geometry in the numerical model by decreasing the diameter of the internal channel from 10 mm to 9.28 mm and add additional losses at the external coupling. The fitting results are shown in Fig. 2(b).

## 4. Scattering parameters characterization

Since circulator can be regarded as the three-isolator combination, for simplicity, the scattering problem of the 3-port acoustic system can be considered as 2-port system one by one, and the corresponding 2-port scattering matrix is given by

$$\begin{vmatrix} B \\ C \end{vmatrix} = |S|_{2\times 2} \begin{vmatrix} A \\ D \end{vmatrix} \tag{A17}$$

where B and C are coefficients describing outgoing waves, A and D are coefficients describing incoming waves. In experiment, we only consider the three harmonics $n = \{-1, 0, 1\}$ that dominate the field in the frequency range of interest. Consequently, the scattering matrix becomes

$$\begin{vmatrix} B^{-1} \\ C^{-1} \\ B \\ C \\ B^{+1} \\ C^{+1} \end{vmatrix} = |S|_{6\times 6} \begin{vmatrix} A^{-1} \\ D^{-1} \\ A \\ D \\ A^{+1} \\ D^{+1} \end{vmatrix} \tag{A18}$$

where $A^{-1}$, $B^{-1}$, $C^{-1}$ and $D^{-1}$ are coefficients related with the -1 order mode, $A^{+1}$, $B^{+1}$, $C^{+1}$ and $D^{+1}$ are coefficients related with the +1 order mode. Since anechoic terminations are connected to ports 2 and 3, the $D$ related coefficients vanish. The transmission from port 1 to port 2 or 3 is defined as

$$S_{21}(S_{31}) = \frac{C}{A} = s_{41}\frac{A^{-1}}{A} + s_{43} + s_{45}\frac{A^{+1}}{A} \tag{A19}$$

where $s_{ij}$ is the element of the scattering matrix. $A^{-1}$ and $A^{+1}$ parasitic signal amplitudes, which are relatively small compared with the excitation signal. Therefore, the transmission is further simplified as

$$S_{21}(S_{31}) = \frac{C}{A} = s_{43} \tag{A20}$$

Similarly, the reflection $S_{11}$ is derived as

$$S_{11} = \frac{B}{A} = s_{33} \tag{A21}$$

## 5. Description of the experimental setup

The coupled resonators were 3D printed using photosensitive resin (modulus 2.65 GPa, density 1130 kg/m³). The thinnest wall is 5 mm thick to ensure good confinement of the acoustic waves. Small holes (diameter 2.5 mm) were drilled in the sample to insert the rods (diameter 2 mm) connecting the plates to the gears. Anechoic terminations were built using conical foam of about 1.5 resonance wavelengths so that absorption coefficients higher than 0.95 were measured at ports 2 and 3 over the working frequency range in the absence of the resonator. To guarantee plane wave propagation, we ensure that the first cutoff frequency ($f_c = c_0/2a \approx 5716$ Hz) of the waveguide remains well above the maximum frequency of interest.

For static case, the scattering parameters were measured by a 15 seconds chirp signal covering the 1000 Hz to 2500 Hz frequency range. The data were acquired using the LMS SCADAS hardware, and LMS Test.Lab analysis software. For the time modulated case, the scattering parameters were measured by three sets of two microphones (BSWA MPA416, 1/4 inch) separated by $s = 50$ mm to discriminate between incident and reflected waves on each port. Here, the microphone spacing is less than the shortest half wavelength of interest. The loudspeaker delivered 3 seconds sinusoidal signals with different excitation frequency each time. The scattering parameters is obtained around resonance frequency ($f_r \pm 50$ Hz) by 2-Hz step while a 5-Hz step was used away from the resonance frequency ($|f - f_r| > 50$ Hz). The resonance frequency was measured by fixing the output power of the motor at a reasonable value while varying the excitation frequency. The frequency resolution was fixed to 0.098 Hz with sample frequency 6400 Hz. Furthermore, the modulation frequency was obtained by calculating the difference value between excitation frequency and +1/-1 Floquet harmonics.

[1]  J. W. S. Rayleigh, *The Theory of Sound* (Dover Publications, USA, 1945).

[2]  P. M. Morse and K. U. Ingard, *Theoretical Acoustics* (McGraw-Hill, New York, 1968).

[3]  R. Fleury, D. Sounas, M. R. Haberman, and A. Alù, Nonreciprocal Acoustics, Acous. Today **11**, 14 (2015).

[4]  H. Nassar, B. Yousefzadeh, R. Fleury, M. Ruzzene, A. Alù, C. Daraio, A. N. Norris, G. Huang, and M. R. Haberman, Nonreciprocity in acoustic and elastic materials, Nat. Rev. Mater. **5**, 667 (2020).

[5]  C. Kittel, Interaction of spin waves and ultrasonic waves in ferromagnetic crystals, Phys. Rev. **110**, 836 (1958).

[6]  B. Liang, X. S. Guo, J. Tu, D. Zhang, and J. C. Cheng, An acoustic rectifier, Nat. Mater. **9**, 989 (2010).

[7]  B. Liang, B. Yuan, and J. C. Cheng, Acoustic diode: rectification of acoustic energy flux in one-dimensional systems, Phys. Rev. Lett. **103**, 104301 (2009).

[8]  L. Shao, W. Mao, S. Maity, N. Sinclair, Y. Hu, L. Yang, and M. Lončar, Non-reciprocal transmission of microwave acoustic waves in nonlinear parity–time symmetric resonators, Nat. Electron. **3**, 267 (2020).

[9]  X. Guo, H. Lissek, and R. Fleury, Observation of non-reciprocal harmonic conversion in real sounds, Commun. Phys. **6**, 93 (2023).

[10]  R. Fleury, D. L. Sounas, C. F. Sieck, M. R. Haberman, and A. Alù, Sound Isolation and Giant Linear Nonreciprocity in a Compact Acoustic Circulator, Science **343**, 516 (2014).

[11] Y. Ding, Y. Peng, Y. Zhu, X. Fan, J. Yang, B. Liang, X. Zhu, X. Wan, and J. Cheng, Experimental Demonstration of Acoustic Chern Insulators, Phys. Rev. Lett. **122**, 014302 (2019).

[12] Y. Zhu, L. Cao, A. Merkel, S.-W. Fan, B. Vincent, and B. Assouar, Janus acoustic metascreen with nonreciprocal and reconfigurable phase modulations, Nat. Commun. **12**, 7089 (2021).

[13] A. B. Khanikaev, R. Fleury, S. H. Mousavi, and A. Alù, Topologically robust sound propagation in an angular-momentum-biased graphene-like resonator lattice, Nat. Commun. **6**, 8260 (2015).

[14] Y. Shi, Z. Yu, and S. Fan, Limitations of nonlinear optical isolators due to dynamic reciprocity, Nat. Photonics **9**, 388 (2015).

[15] T. Pedergnana, A. Faure-Beaulieu, R. Fleury, and N. Noiray, Loss-compensated non-reciprocal scattering based on synchronization, Nat. Commun. **15**, 7436 (2024).

[16] C. Shen, J. Li, Z. Jia, Y. Xie, and S. A. Cummer, Nonreciprocal acoustic transmission in cascaded resonators via spatiotemporal modulation, Phys. Rev. B **99**, 134306 (2019).

[17] J. Li, C. Shen, X. Zhu, Y. Xie, and S. A. Cummer, Nonreciprocal sound propagation in space-time modulated media, Phys. Rev. B **99**, 144311 (2019).

[18] X. Zhu, J. Li, C. Shen, X. Peng, A. Song, L. Li, and S. A. Cummer, Non-reciprocal acoustic transmission via space-time modulated membranes, Appl. Phys. Lett. **116**, 034101 (2020).

[19] T. T. Koutserimpas, A. Alù, and R. Fleury, Parametric amplification and

bidirectional invisibility in PT-symmetric time-Floquet systems, Phys. Rev. A **97**, 013839 (2018).

[20] X. Wen, X. Zhu, A. Fan, W. Y. Tam, J. Zhu, H. W. Wu, F. Lemoult, M. Fink, and J. Li, Unidirectional amplification with acoustic non-Hermitian space−time varying metamaterial, Commun. Phys. **5**, 18 (2022).

[21] M. Malléjac and R. Fleury, Scattering from Time-Modulated Transmission-Line Loads: Theory and Experiments in Acoustics, Phys. Rev. Appl. **19**, 064012 (2023).

[22] J. Li, X. Zhu, C. Shen, X. Peng, and S. A. Cummer, Transfer matrix method for the analysis of space-time-modulated media and systems, Phys. Rev. B **100**, 144311 (2019).

[23] R. Fleury, A. B. Khanikaev, and A. Alù, Floquet topological insulators for sound, Nat. Commun. **7**, 11744 (2016).

[24] T. T. Koutserimpas and R. Fleury, Zero refractive index in time-Floquet acoustic metamaterials, J. Appl. Phys. **123**, 091709 (2018).

[25] C. Shen, X. Zhu, J. Li, and S. A. Cummer, Nonreciprocal acoustic transmission in space-time modulated coupled resonators, Physical Review B **100**, 054302 (2019).

[26] Z. Chen, Z. Li, J. Weng, B. Liang, Y. Lu, J. Cheng, and A. Alù, Sound non-reciprocity based on synthetic magnetism, Sci. Bull. **68**, 2164 (2023).

[27] R. Fleury, D. L. Sounas, and A. Alù, Subwavelength ultrasonic circulator based on spatiotemporal modulation, Phys. Rev. B **91**, 174306 (2015).

[28] A. Darabi, X. Ni, M. Leamy, and A. Alù, Reconfigurable Floquet elastodynamic topological insulator based on synthetic angular momentum bias, Sci. Adv. **6**, eaba8656 (2020).

[29] M. Malléjac and R. Fleury, Experimental realization of an active time-modulated acoustic circulator, *arXiv:2409.04251*, (2024).

[30] H. A. Haus, *Waves and Fields in Optoelectronics* (Prentice Hall 1984).